\documentclass[journal=jacsat,manuscript=article]{achemso}

\usepackage[version=3]{mhchem} % Formula subscripts using \ce{}
\usepackage[utf8]{inputenc}
\usepackage{color}
\usepackage{tabularray}
\usepackage{amsmath} 
\usepackage{enumitem}
\usepackage{graphicx}
\usepackage{epsfig}
\usepackage{booktabs}
\usepackage{multirow}
\usepackage{float}

\usepackage{tabularx}
\usepackage{booktabs} 
\usepackage{adjustbox}
\usepackage{hyperref}
\usepackage{siunitx}
\usepackage{xcolor}
\author{Suryanarayanan Balaji}
\affiliation[cheme]
{Department of Chemical Engineering, Carnegie Mellon University, Pittsburgh PA, USA 15213}
\author{Rishikesh Magar}
\affiliation[meche]
{Department of Mechanical Engineering, Carnegie Mellon University, Pittsburgh PA, USA 15213}
\author{Yayati Jadhav}
\affiliation[meche]
{Department of Mechanical Engineering, Carnegie Mellon University, Pittsburgh PA, USA 15213}
\author{Amir Barati Farimani}
\email{barati@cmu.edu}
\affiliation[meche]
{Department of Mechanical Engineering, Carnegie Mellon University, Pittsburgh PA, USA 15213}
\alsoaffiliation[cheme]
{Department of Chemical Engineering, Carnegie Mellon University, Pittsburgh PA, USA 15213}

\title[An \textsf{achemso} demo]
  {GPT-MolBERTa: GPT Molecular Features Language Model for molecular property prediction}

\abbreviations{IR,NMR,UV}
\keywords{American Chemical Society, \LaTeX}

\begin{document}

%%%%%%%%%%%%%%%%%%%%%%%%%%%%%%%%%%%%%%%%%%%%%%%%%%%%%%%%%%%%%%%%%%%%%
%% The "tocentry" environment can be used to create an entry for the
%% graphical table of contents. It is given here as some journals
%% require that it is printed as part of the abstract page. It will
%% be automatically moved as appropriate.
%%%%%%%%%%%%%%%%%%%%%%%%%%%%%%%%%%%%%%%%%%%%%%%%%%%%%%%%%%%%%%%%%%%%%
% \begin{tocentry}

% Some journals require a graphical entry for the Table of Contents.
% This should be laid out ``print ready'' so that the sizing of the
% text is correct.

% Inside the \texttt{tocentry} environment, the font used is Helvetica
% 8\,pt, as required by \emph{Journal of the American Chemical
% Society}.

% The surrounding frame is 9\,cm by 3.5\,cm, which is the maximum
% permitted for  \emph{Journal of the American Chemical Society}
% graphical table of content entries. The box will not resize if the
% content is too big: instead it will overflow the edge of the box.

% This box and the associated title will always be printed on a
% separate page at the end of the document.

% \end{tocentry}
%%%%%%%%%%%%%%%%%%%%%%%%%%%%%%%%%%%%%%%%%%%%%%%%%%%%%%%%%%%%%%%%%%%%%
%% The abstract environment will automatically gobble the contents
%% if an abstract is not used by the target journal.
%%%%%%%%%%%%%%%%%%%%%%%%%%%%%%%%%%%%%%%%%%%%%%%%%%%%%%%%%%%%%%%%%%%%%
\begin{abstract}
 
  \noindent With the emergence of Transformer architectures and their powerful understanding of textual data, a new horizon has opened up to predict the molecular properties based on text description. While SMILES are the most common form of representation, they are lacking robustness, rich information and canonicity, which limit their effectiveness in becoming generalizable representations. Here, we present GPT-MolBERTa, a self-supervised large language model (LLM) which uses detailed textual descriptions of molecules to predict their properties. A text based description of 326000 molecules were collected using ChatGPT and used to train LLM to learn the representation of molecules. To predict the properties for the downstream tasks, both BERT and RoBERTa models were used in the finetuning stage. Experiments show that GPT-MolBERTa performs well on various molecule property benchmarks, and approaching state of the art performance in regression tasks. Additionally, further analysis of the attention mechanisms show that GPT-MolBERTa is able to pick up important information from the input textual data, displaying the interpretability of the model.

\end{abstract}

%%%%%%%%%%%%%%%%%%%%%%%%%%%%%%%%%%%%%%%%%%%%%%%%%%%%%%%%%%%%%%%%%%%%%
%% Start the main part of the manuscript here.
%%%%%%%%%%%%%%%%%%%%%%%%%%%%%%%%%%%%%%%%%%%%%%%%%%%%%%%%%%%%%%%%%%%%%
\section{Introduction}
Molecular property prediction is vital for drug discovery, guiding compound selection, evaluation, and generation \cite{JieShen2020, Bartok2013, Huang2016}, however, experiments to determine molecular properties can be very expensive and time consuming. Computational techniques such as machine learning (ML) can be an excellent approach to predict the properties since they are fast and can directly map the molecules to their properties. To develop accurate molecular property prediction models, an important factor to consider is the molecular representation, which involves encoding chemical compounds for computational analysis. \cite{LaurianneDavid2020}. Different methods for representing molecules include SMILES\cite{Weininger1987}, graph-based representations\cite{Duvenaud2015}, and molecular fingerprints\cite{rogers2010extended,capecchi2020one}. Typically, Graph Neural Networks (GNNs)\cite{Gilmer2017, Dejun2021, KevinYang2019, Gasteiger2022, Schutt2018, ChengQiang2019, Karamad2020, Hoon2023} show superior performance, capturing detailed geometric and atomic neighborhood information. However, their interpretability can be limited. In contrast, SMILES, a string-based representation, stands out for its simplicity and adaptability\cite{Wigh2022, Cheng2023, Zhang2022, Sachdev2019, Guo2022, Rogers2010, Gomez2018}. The inherent string based nature of SMILES makes them well-suited for transformer-like architectures, and this is further enhanced by the availability of extensive databases for training\cite{PubChemSize,Gaulton2016}

Transformers, initially developed for natural language processing tasks\cite{Vaswani2017, Tianyang2022, Kalyan2021, Jianlin2022, Katharopoulos2020, Devlin2019, Liu2019}, are now increasingly being exploited in molecular ML. Transformer models that utilize SMILES as inputs have emerged for molecular property prediction and generation\cite{Fabian2020, Brown2019, Irwin2022, Sterling2015, Bjerrum2018, Chithrananda2020, Ahmad2022, Ross2022, Yuksel2023, Krenn2019, Born2023, Zhonglin2023, Guntuboina2023peptidebert, Huang2023materials, Xu2023transpolymer}. Their sequential data processing capability, combined with the inherent attention mechanism, provides some level of interpretability. While SMILES play a significant role in the areas of molecular property prediction, modeling, and design \cite{Gilmer2017, Wang2019, Rishi2021, Chmiela2018, Altae2017, wang2020bio}, they have inherent limitations. SMILES are non-canonical in nature, where a single SMILES string could represent multiple molecules \cite{Wigh2022}. Additionally, SMILES fail to encode the topographical information of the molecule such as 3D structure and stereochemistry, limiting the performance of many machine learning models \cite{Yuyang2022}. Considering these challenges, it prompts the question: Can we develop a representation that maintains the simplicity of SMILES, yet embeds explicit details about a molecule's structural attributes and potentially incorporates geometric insights? Developing such a representation could enhance the performance of transformer models in molecular property prediction.

Models like MatBERT \cite{Trewartha2022} and MatSciBERT \cite{Gupta2022}, pretrained on material science tasks, show promise in property predictions. These results imply that domain-specific pretraining might be more beneficial than just using larger transformer models. However, using molecular domain papers for pretraining might not provide specific information. An alternative is generating unique text descriptions for SMILES molecules using large language models, with models like ChatGPT \cite{Qian2023, Ock2023} showing potential in this area.

Textual descriptions give a broader look at molecules, covering everything from basic atomic details to complex geometric information and interactions. SMILES notation is good at providing an overall view of the molecule, however, textual description provides more details and is more comprehensive. For example, for water molecule, hundred pages of information is available (for example, this text is taken from water molecule from Wikipedia page: {\it"Water is an inorganic compound with the chemical formula H2O. It is a transparent, tasteless, odorless, and nearly colorless chemical substance, and it is the main constituent of Earth's hydrosphere and the fluids of all known living organisms, the bond angle between ..."}). The depth provided by text might help improve how we model and predict molecular properties, blending the best of both SMILES and geometry based graphs representations.

\begin{figure}[htbp]
\includegraphics[height = 16 cm, width = 15 cm]{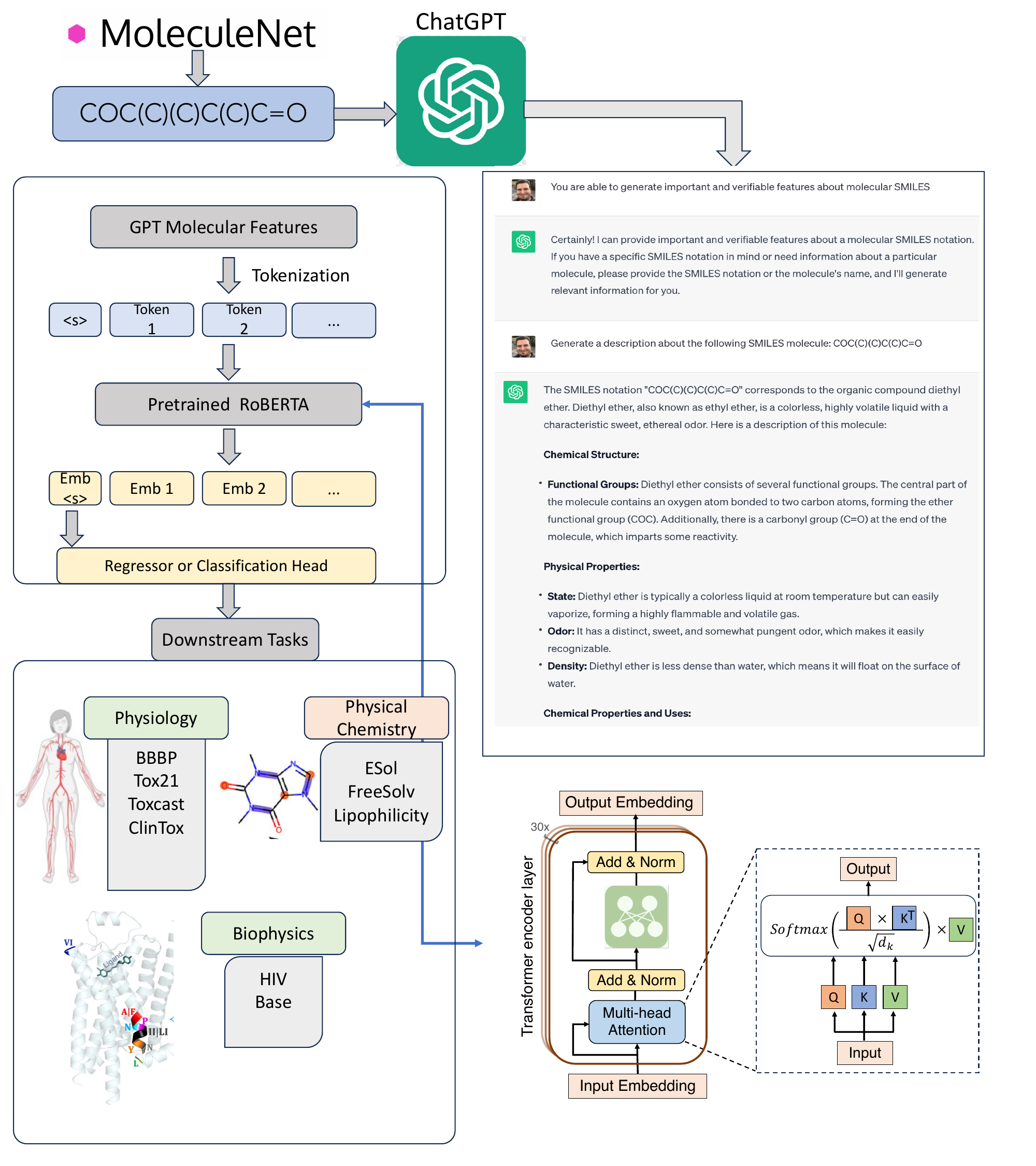}
\caption{Overview of GPT-MolBERTa. SMILES strings are sent to ChatGPT, which generates rich textual descriptions consisting of information about functional groups, molecular weight, density, and other properties. These descriptions are then used to pretrain a RoBERTa model. The model is then fine-tuned on MoleculeNet datasets, with the addition of a classification/regression head to the first token embeddings.}
\label{Figure 1: GPT-MolBERTa procedure}
\end{figure}

In this paper, we introduce GPT-MolBERTa (GPT Molecule-RoBERTa), a chemical language model that leverages molecular text descriptions as inputs for downstream molecular property prediction tasks. The text descriptions were generated through the use of generative large language models, in this case OpenAI's \cite{openai2023} ChatGPT. Figure~\ref{Figure 1: GPT-MolBERTa procedure} provides an overview of the methodology used in this paper. The initial step involves sending SMILES strings into ChatGPT, where they are used to generate rich textual descriptions through the use of an optimized prompt. These descriptions include details about functional groups, shape, and chemical properties and are consolidated into a text corpus. This corpus is subsequently used as input for a RoBERTa model, upon which a classification/regression head was added for molecular property prediction. The model was pretrained on the text descriptions of approximately 326,000 molecules sourced from MoleculeNet \cite{Zhenqin2018} and evaluated on MoleculeNet's benchmark datasets.

Notably, GPT-MolBERTa demonstrates strong overall performance, approaching state-of-the-art levels in regression tasks. The promising aspect is that the model's pretraining was conducted with only around 300,000 molecules, a significantly smaller dataset compared to other models that use millions of molecules. This finding suggests that pretraining with a comparable number of text descriptions holds the potential to improve molecular property prediction, offering exciting possibilities for future research and applications.

\section{Methods}

GPT-MolBERTa consists of two sections, namely the data generation and the model pretraining and finetuning (Figure \ref{Figure 1: GPT-MolBERTa procedure}). Text descriptions of each molecule were first generated through the use of ChatGPT. These were used to train a self-supervised transformer based encoder model, which was used to extract high dimensional vector representation of the molecular descriptions. This pretrained model was then finetuned on the downstream molecular property prediction benchmarks from MoleculeNet, through the addition of a regression or classification head onto the architecture. Both pretrained LLM models of BERT and RoBERTa were explored to compare their efficiency in learning. The complete model implementation was executed using HuggingFace \cite{Wolf2020} within the PyTorch framework.

\subsection{Dataset Generation and Curation}
We generate the textual descriptions of SMILES molecules through the use of ChatGPT-3.5 \cite{openai2023}. We conduct prompt engineering to obtain the specific prompt which will generate meaningful data. Information about the molecular structures, atomic weights, functional groups were all obtained from the input. Based on our best practices and prompt engineering, we found out that priming the ChatGPT with the prompt {\it " You are able to generate important and verifiable features about molecular SMILES"} and then prompting {\it " Generate a description about the following SMILES molecule ..."} results in the best description of molecules. (Figure \ref{Figure 1: GPT-MolBERTa procedure}) These descriptions were subsequently added to the original dataset, with insignificant responses (comprising fewer than 100 tokens) being excluded. Approximately 326000 text description of molecules were obtained from 14 datasets present in MoleculeNet \cite{Zhenqin2018}. 

\subsection{Tokenization}

Tokenization is a fundamental step in text processing \cite{Webster1992} which involves breaking input text into indivisible units. RoBERTa employs Byte Level Byte-Pair Encoding, ensuring no unknown tokens \cite{Radford2019}. We followed RoBERTa protocols when tokenizing inputs for our model. After tokenization, the tokens are further processed into input embeddings. Positional encoding is added, embedding token positions in the sequence for tasks like prediction and generation.

\subsection{Transformer Model}

The GPT MolBerta's transformer encoder consists of multiple stacked layers each consisting of multi-head self attention layers followed by feed-forward networks. The input data is tokenized and positionally encoded before transformed into embeddings. These embeddings will be sent to the self-attention layers which will extract context information of the input data followed by a feed-forward layer to obtain the final representations. 

The embeddings are subsequently passed on to the attention layers. Both BERT \cite{Devlin2019} and RoBERTa \cite{Liu2019} models use the multi-head scaled dot product attention mechanism, allowing for parallel processing of input tokens. In the self-attention mechanism, each token in the sequence is projected into its corresponding query, key and value vector (Q, K and V) through the use of learnable weight matrices ($W_Q$, $W_K$ and $W_V$). Given that $d_k$ is the dimension of Q and K, the scaled dot-product attention A for a single head is calculated by the following equation \cite{Vaswani2017}.
\[
\begin{gathered}
\text{Attention (Q, K, V)} = \text{Softmax}\left(\frac{QK^T}{\sqrt{d_k}}\right)V
\end{gathered}
\]

Multi-head attention conducts these calculations in parallel across different heads, allowing for the model to jointly attend to information from different representation subspaces at different positions \cite{Vaswani2017}. These outputs will be concatenated and projected into the output embedding with the same size as the input embedding. The operation is shown below.
\[
\begin{gathered}
\text{ MultiHead (Q, K, V)} = \text{ Concat}(Head_1, \ldots, Head_h)W^O \\
\text{ where } Head_i = \text{ Attention}(QW^Q_i, KW^K_i, VW^V_i)
\end{gathered}
\]

\noindent The outputs would then be sent to a feed-forward network to transform the attention-derived features. Layer normalization and residual connections are employed throughout the encoder layer to enhance training stability and convergence speed. Multiple encoder layers are used to improve the quality of the embeddings obtained.

\subsection{Pretraining}
Pretraining a large language model involves training the model on an extensive corpus of data before finetuning it on a specific downstream task. In this study, we pretrained both the BERT and RoBERTa tokenizers using a molecular text corpus to extract their specific vocabularies. While both BERT and RoBERTa have already been pretrained on the BooksCorpus \cite{Zhu2015} and English Wikipedia, we opted to train our own tokenizer as it would be tuned specifically to the vocabulary in our dataset, allowing for better identification of tokens. In the case of the RoBERTa tokenizer, the special tokens were post processed to ensure parity with the existing RoBERTa vocabulary. 

Following tokenization, the models were pretrained in a self-supervised \cite{Magar2022, Chen2020, Zbontar2021} manner using masked language modeling \cite{Devlin2019}, where 15\% of the input tokens were masked and the model would predict the masked tokens using bidirectional context. RoBERTa goes one step further through introducing dynamic token masking, where different tokens in the sequence are masked per epoch. Masked language modeling allows the model to learn meaningful representation of the input data, without the need for labels.

\subsection{Finetuning}
GPT-MolBERTa was finetuned on property prediction tasks from several benchmark datasets present in MoleculeNet, the details of which are given in Table 1. The final molecular property prediction will be conducted through adding a classification or regression head to the embeddings from the first token. Early stopping was also implemented if the validation loss does not show improvement over a specified number of epochs, minimizing the risk of overfitting.

For a given dataset, we first removed all non-canonical SMILES strings. This filtering along with the earlier removal of insignificant responses accounts for 0.14\% of the dataset. We use the binary cross-entropy and root mean squared error (RMSE) as our loss functions and the Adam optimizer for our training. All benchmarks were scaffold split in the ratio of 80/10/10 to match the standards used in MoleculeNet. Scaffold splitting splits molecules according to their Murcko Scaffolds, making the train and test datasets as dissimilar to each other, resulting in a more challenging task. For datasets involving multiple labels, we adopted a consistent methodology: training and validation were conducted for each label using the same model. The subsequent averages were then computed and reported for both training and testing phases. This process was repeated three times per dataset to determine average and standard deviation performance on the test set. Hyper-parameters are shown in S4 of Supplementary Information.

\begin{table}
\centering
\begin{tabular}{l|l|lll}
\toprule
Task (Metric)                             & Dataset       & \# Molecules &  &  \\
\midrule
\multirow{3}{*}{Regression(RMSE)}         & ESOL          & 1128         &  &  \\
                                          & FreeSolv      & 642          &  &  \\
                                          & Lipophilicity & 4200         &  &  \\
\midrule
\multirow{6}{*}{Classification (ROC-AUC)} & HIV           & 41127        &  &  \\
                                          & BACE          & 1513         &  &  \\
                                          & BBBP          & 2039         &  &  \\
                                          & Tox21         & 7831         &  &  \\
                                          & SIDER         & 1427         &  &  \\
                                          & ClinTox       & 1478         &  &  \\
\bottomrule
\end{tabular}
\caption{Datasets from the MoleculeNet used for finetuning tasks. We finetune our model on three regression datasets and 6 classification datasets.}
\end{table}

\section{Results and Discussion}

To evaluate GPT-MolBERTa's effectiveness, we conducted a comprehensive benchmark on various classification and regression tasks from the MoleculeNet datasets. The results are summarized in Table 2, comparing the model's test area under the curve (AUC) to baseline models. The averages and standard deviations from three runs were reported, with two models used for comparison.

\subsection{MoleculeNet Benchmark}

The MoleculeNet dataset tasks are divided into regression and classification categories. For classification, we evaluate six datasets: BBBP, Tox21, ClinTox, HIV, BACE, and SIDER, each highlighting different molecular properties. Additionally, we also consider three regression datasets: ESOL, FreeSOLV, and Lipophilicity.

In terms of classification tasks, our model's performance aligns with other baseline models utilizing string-based representations(Table~\ref{Tb:Class}). Additionally, our results are consistent with some graph neural networks, such as GIN \cite{Xu2019} and GCN \cite{Kipf2017}. Overall, our model demonstrates moderate success across the benchmark classification datasets when compared to both GNN and string-based model baselines. It's noteworthy that GPT-MolBERTa was pretrained on a dataset of 326,000 points, smaller than other baselines that were pretrained on datasets an order of magnitude larger. We believe that pretraining on a more extensive corpus might enhance our framework's downstream performance.

We assessed our model's performance on the MoleculeNet regression tasks, with results presented in Table~\ref{Tb:Reg}. This table lists the root mean squared error (RMSE) for each regression dataset. While GPT-MolBERTa posts solid results in classification, it truly stands out in regression tasks. Specifically, it outperforms other GNN models and baseline models that use string-based representations, especially in the FreeSolv and ESOL datasets. For the Lipophilicity dataset, GPT-MolBERTa's performance aligns closely with other baselines. Notably, our model registers a performance gain of 5.88\% over the top-performing baseline, MolBERT, for the FreeSolv dataset, and an 11.32\% improvement for the ESOL dataset.

\begin{table}[htb!]
\centering

\resizebox{\textwidth}{!}{
\begin{tabular}{l|llllll}
\toprule
Models & BBBP & Tox 21 & ClinTox & HIV & BACE & SIDER \\
\midrule
GCN\cite{Kipf2017} & 71.9 ± 0.9 & 70.9 ± 2.6 & 62.5 ± 2.8 & 74.0 ± 3.0 & 71.6 ± 2.0 & 53.6 ± 3.2 \\
GIN\cite{Xu2019} & 65.8 ± 4.5 & 74.0 ± 0.8 & 58.0 ± 4.4 & 75.3 ± 1.9 & 70.1 ± 5.4 & 57.3 ± 1.6 \\
\midrule
SchNet\cite{Schutt2018} & 84.8 ± 2.2 & 77.2 ± 2.3 & 71.5 ± 3.7 & 70.2 ± 3.4 & 76.6 ± 1.1 & 53.9 ± 3.7 \\
MGCN\cite{ChengQiang2019} & \textit{85.0 ± 6.4} & 70.7 ± 1.6 & 63.4 ± 4.2 & 73.8 ± 1.6 & 73.4 ± 3.0 & 55.2 ± 1.8 \\
D-MPNN\cite{KevinYang2019} & 71.2 ± 3.8 & 68.9 ± 1.3 & 90.5 ± 5.3 & 75.0 ± 2.1 & 85.3 ± 5.3 & 63.2 ± 2.3 \\
Hu et al.\cite{Hu2020} & 70.8 ± 1.5 & 78.7 ± 0.4 & 78.9 ± 2.4 & 80.2 ± 0.9 & 85.9 ± 0.8 & 65.2 ± 0.9 \\
MolCLR-GCN\cite{Yuyang2022} & 73.8 ± 0.2 & 74.7 ± 0.8 & 86.7 ± 1.0 & 77.8 ± 0.5 & 78.8 ± 0.5 & 66.9 ± 1.2 \\
MolCLR-GIN\cite{Yuyang2022} & 73.6 ± 0.5 & \textit{79.8 ± 0.7} & \textit{93.2 ± 1.7} & \textit{80.6 ± 1.1} & \textit{89.0 ± 0.3} & \textit{68.0 ± 1.1} \\
\midrule
MolBERT\cite{Fabian2020} & 76.2 ± 0.0 & - & - & 78.3 ± 0.0 & \textbf{86.6 ± 0.0} & - \\
ChemBERTa-2\cite{Ahmad2022} & 72.8 ± 0.0 & - & - & 62.2 ± 0.0 & 79.9 ± 0.0 & - \\
CLM\cite{Born2023} & \textbf{91.5 ± 0.0} & \textbf{79.5 ± 0.0} & - & \textbf{81.3 ± 0.0} & 86.1 ± 0.0 & 61.9 ± 0.0 \\
SELFormer\cite{Yuksel2023} & 90.2 ± 0.0 & 65.3 ± 0.0 & - & 68.1 ± 0.0 & 83.2 ± 0.0 & \textbf{74.5 ± 0.0} \\
GPT-MolBERTa & 74.1 ± 0.15 & 65.9 ± 0.06 & \textbf{49.7 ± 0.12} & 75.5 ± 1.29 & 73.4 ± 0.47 & 58.5 ± 0.35 \\
\bottomrule

\end{tabular}
}

\caption{Classification Benchmarks on MoleculeNet. We benchmark the model against standard GNN baseline as well as transformer baselines. The evaluation metric used for classification tasks is ROC-AUC. The best performing result among the string representation based approaches has been shown in boldface and the best performing GNN result has been italicized.}
\label{Tb:Class}
\end{table}

\begin{table}[H]
\centering
\resizebox{\textwidth}{!}{
\begin{tabular}{l|lll}
\toprule
Models & FreeSolv & ESOL & Lipophilicity \\
\midrule
GCN\cite{Kipf2017} & 2.87 ± 0.14 & 1.43 ± 0.05 & 0.85 ± 0.08 \\
GIN\cite{Xu2019} & 2.76 ± 0.18 & 1.45 ± 0.02 & 0.85 ± 0.07 \\
\midrule
SchNet\cite{Schutt2018} & 3.22 ± 0.76 & 1.05 ± 0.06 & 0.91 ± 0.10 \\
MGCN\cite{ChengQiang2019} & 3.35 ± 0.01 & 1.27 ± 0.15 & 1.11 ± 0.04 \\
D-MPNN\cite{KevinYang2019} & \textit{2.18 ± 0.91} & \textit{0.98 ± 0.26} & \textit{0.65 ± 0.05} \\
Hu et al.\cite{Hu2020} & 2.83 ± 0.12 & 1.22 ± 0.02 & 0.74 ± 0.00 \\
MolCLR-GCN\cite{Yuyang2022} & 2.39 ± 0.14 & 1.16 ± 0.00 & 0.78 ± 0.01 \\
MolCLR-GIN\cite{Yuyang2022} & 2.20 ± 0.20 & 1.11 ± 0.01 & \textit{0.65 ± 0.08} \\
\hline
MolBERT\cite{Fabian2020} & 0.948 ± 0.33 & 0.531 ± 0.04 & \textbf{0.561 ± 0.03} \\
ChemBERTa-2\cite{Ahmad2022} & - & - & 0.798 ± 0.00 \\
ChemFormer\cite{Irwin2022} & 1.23 ± 0.00 & 0.633 ± 0.00 & 0.598 ± 0.00 \\
SELFormer\cite{Yuksel2023} & 2.797 ± 0.00 & 0.682 ± 0.00 & 0.735 ± 0.00 \\
GPT-MolBERTa & \textbf{0.896 ± 0.02} & \textbf{0.477 ± 0.01} & 0.758 ± 0.01 \\
\bottomrule

\end{tabular}
}
\caption{Regression Benchmarks on MoleculeNet. We benchmark the model against standard GNN baseline as well as transformer baselines. The evaluation metric used for regression tasks is RMSE. The best performing result among the string representation based approaches has been shown in boldface and the best performing GNN result has been italicized.}
\label{Tb:Reg}
\end{table}

\subsection{Effect of the Transformer Encoder}
After observing GPT-MolBERTa's performance, we aim to assess the generalizability of our framework. For comparison, we trained a BERT encoder with the same dataset. We tokenized the input using BERT's Word Piece \cite{Nayak2020} Tokenizer and kept the model architecture identical to that of RoBERTa.

With BERT, we observe a similar trend in performance across the MoleculeNet benchmarks. It shows especially strong performance in regression tasks, while aligning with other baseline models utilizing string-based representations for the classification tasks. The model comparisons are shown in Table~\ref{Tb:RoBERTa vs BERT} below.

\begin{table}[htb!]
\centering
\resizebox{\textwidth}{!}{
\begin{tabular}{l|lll}
\toprule
Dataset & BERT & RoBERTa & Change (\%) \\
\midrule
BBBP & 71.3 ± 1.79 & 74.1 ± 0.15 & 3.83 \\
BACE & 74.4 ± 1.53 & 73.4 ± 0.47 & -1.34 \\
ClinTox & 49.6 ± 0.15 & 49.7 ± 0.12 & 0.07 \\
SIDER & 56.7 ± 0.70 & 58.5 ± 0.35 & 3.23 \\
Tox21 & 63.4 ± 0.85 & 65.9 ± 0.06 & 3.94 \\
HIV & 70.6 ± 1.38 & 75.5 ± 1.29 & 7.04 \\
\hline
FreeSolv & 1.006 ± 0.051 & 0.896 ± 0.023 & -10.90 \\
ESOL & 0.531 ± 0.040 & 0.477 ± 0.007 & -10.23 \\
Lipophilicity & 0.810 ± 0.013 & 0.758 ± 0.008 & -6.42 \\
\bottomrule

\end{tabular}
}
\caption{Performance comparison between different transformer encoders. The table presents a performance comparison between different transformer encoders, specifically BERT and RoBERTa, in capturing essential molecular representations. The \% Change column represents the relative improvement of RoBERTa over BERT. Positive values indicate improved performance for classification tasks, while negative values signify better performance in regression tasks.}
\label{Tb:RoBERTa vs BERT}
\end{table}

From Table~\ref{Tb:RoBERTa vs BERT}, it is observed that RoBERTa consistently outperforms BERT in both classification and regression tasks, demonstrating up to 7.04\% and 10.23\% in HIV and ESOL datasets respectively. This suggests that the learned representations exhibit strong generalizability, as the difference in model performance is about 10\%, hinting at the potential for even better performance with more advanced models.

\subsection{Effect of Pretraining}
\begin{figure}[htb!]
    \centering
    \includegraphics[width=1.0\textwidth]{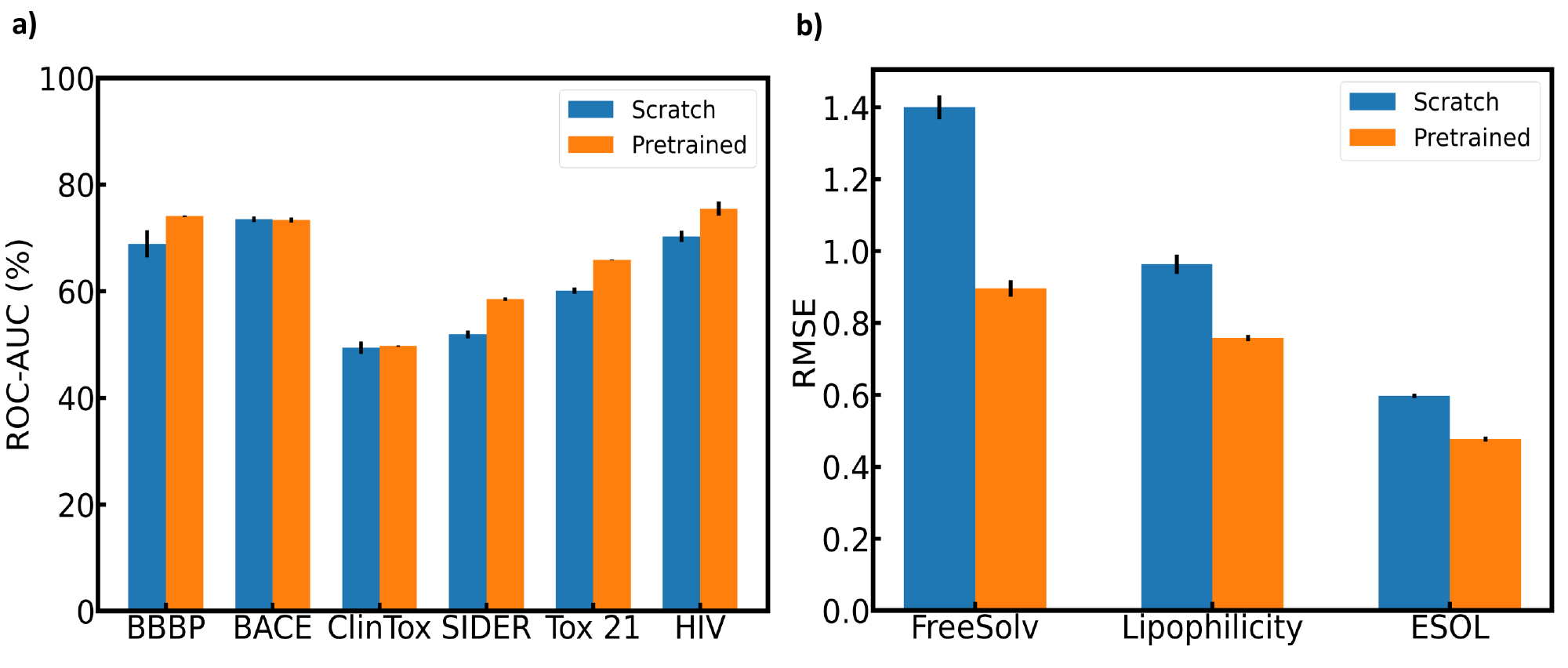}  % Adjust width as needed
    \caption{Effect of Pretraining on GPT-MolBERTa with (a) Classification tasks and (b) Regression tasks. The comparison between the pretrained model and the model trained from scratch is demonstrated for each dataset.}
    \label{Fig:ScratchvsPretrain}
\end{figure}

To evaluate the benefits of pretraining, we compared the performance of two models: one that was trained from scratch and another that was pretrained using Masked Language Modeling. As depicted in Figure~\ref{Fig:ScratchvsPretrain}, there's a clear advantage to pretraining — it leads to a noticeable improvement in property prediction accuracy. This suggests that GPT-MolBERTa can effectively utilize unlabeled data to craft representations that are both meaningful and applicable to molecular property prediction tasks. An added benefit is that pretrained models already have a basic grasp of chemistry. Researchers can further fine-tune these models for specific tasks, combining general and specialized knowledge. 

\subsection{Understanding the Representations}

\begin{figure}[H]
    \centering
    \includegraphics[width=1.0\linewidth]{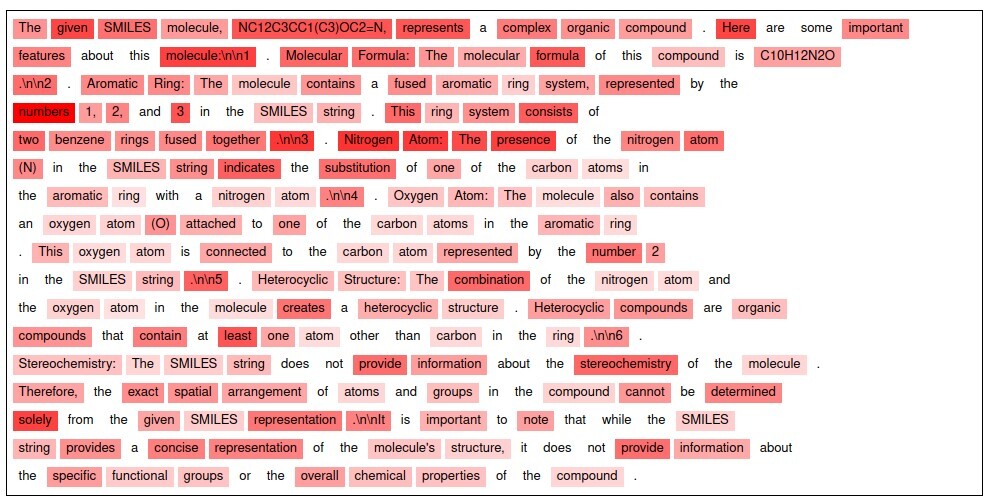}
    \caption{A sample attention map from the model. Given a sample description, it highlights the sections of the descriptions according to its attention scores, showing how the model focuses on specific aspects of the descriptions.}
    \label{fig:Attention}
\end{figure}

Figure~\ref{fig:Attention} displays the attention visualization of the RoBERTa model \cite{Ala2023}. It reveals that the model particularly focuses on certain parts of the description, such as the SMILES string, as well as specific information like atom type and properties. Taking the molecule 'NC12C3CC1(C3)OC2=N' as an example, the attention mechanism underscores terms like "Nitrogen", "stereochemistry", "Aromatic Ring", "rings", "fused", and "heterocyclic". While the SMILES representation can encapsulate some of this data, textual descriptions add an interpretable dimension by assigning word attributes to elements, like specifying "benzene rings". This added interpretability is a significant advantage of our model. This added interpretability is a significant advantage of our model. This increased clarity is a notable benefit of our model. By using specific terms like "benzene rings," the model provides a clearer picture of the molecule's structure and properties. This method offers a balance between a detailed representation and easily understandable information, making it useful for to interpret important characteristic of molecules leveraged by the model for final property prediction.

\begin{figure}[htb!]
    \centering
    \includegraphics[width=1.0\textwidth]{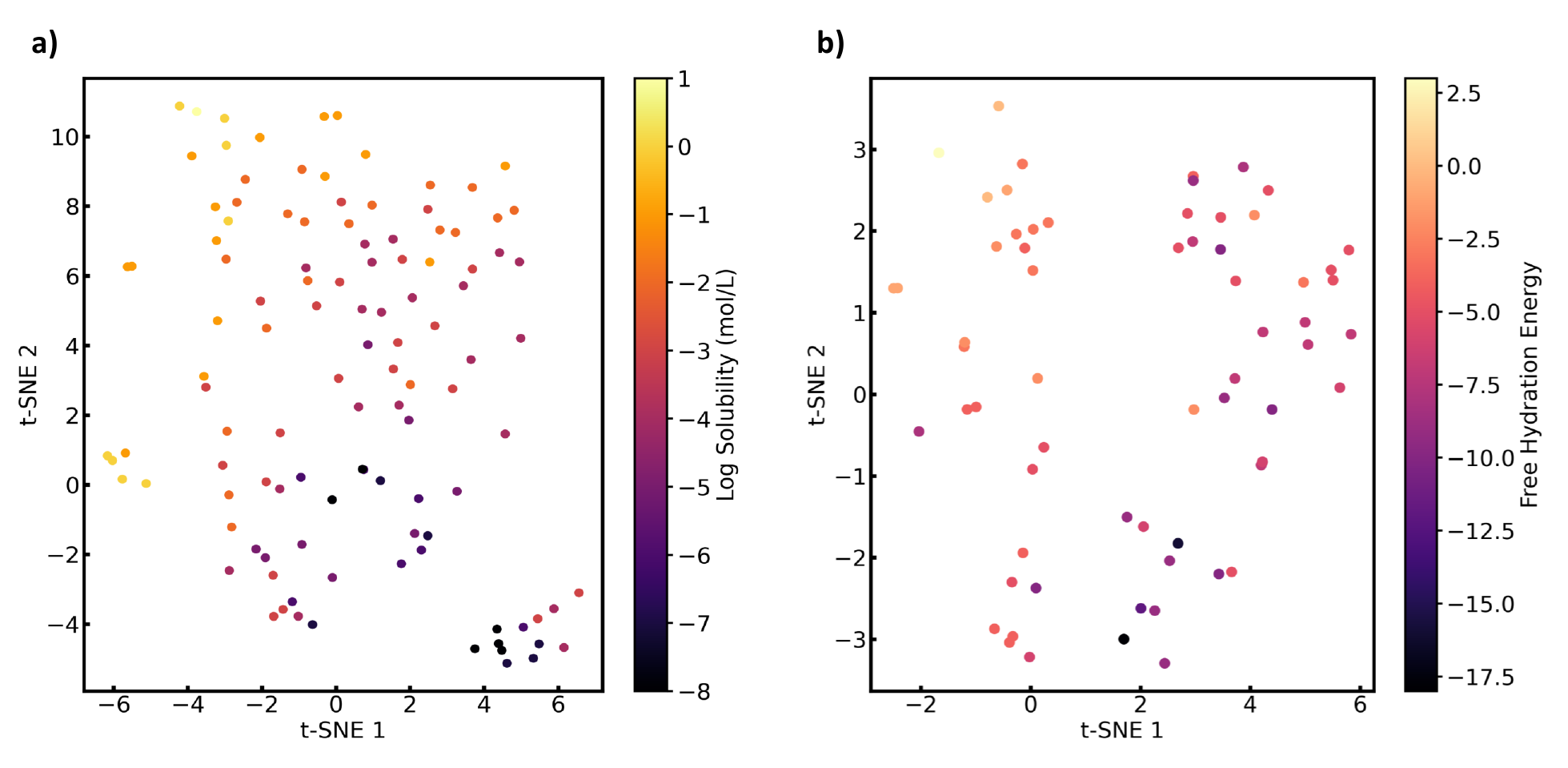} 
    \caption{t-SNE Embeddings of the First Token of GPT-MolBERTa for a) ESOL and b) FreeSolv datasets: Each point in this plot represents log solvation energy for ESOL and free hydration energy for FreeSolv.}
    \label{fig:t-SNE}
\end{figure}

To further delve into the representations learned by GPT-MolBERTa, we employed dimension reduction through t-SNE embedding. We applied the t-SNE algorithm to map the test sets of both the ESOL and FreeSolv datasets, as they showcased the best performance among the models. The resulting visualizations are presented in Figure~\ref{fig:t-SNE}.
Upon closer inspection of these visualizations, an interesting pattern emerges. GPT-MolBERTa demonstrates its ability to effectively cluster labels, where labels exhibiting more negative values are clustered towards the bottom-right, and the more positive values are clustered towards the top-right, observed for both the ESOL and FreeSolv datasets. This observation underscores GPT-MolBERTa's capacity to extract meaningful and informative features from the input data, highlighting its practicality and potential for molecular discovery and property prediction.

\section{Conclusion}
In this work, we introduce GPT-MolBERTa framework, that harnesses text descriptions of molecules to train large-scale language models for molecular property prediction. We successfully demonstrate the viability of our strategy by benchmarking the framework on MoleculeNet. GPT-MolBERTa's ability to represent molecules showed consistent performance across a wide variety of chemicals, suggesting it can generalize well even with limited data. We believe that the performance of the model access to more extensive data, similar to the approach used by more established models like ChemBERTa. 

A distinct feature of our model is an added layer of interpretability by leveraging the attention mechanism. The attention visualizations highlight which parts of the molecular description the model views as most important, offering clearer insights into its decision-making process. Furthermore, we can also look into advanced techniques, such as contrastive learning, as avenues to improve the model's performance. With further refinement, we believe this model can play a pivotal role in applications like drug discovery.

\begin{acknowledgement}

The authors would like to express gratitude to Janghoon Ock for many insightful discussions regarding Transformer-based models.

\end{acknowledgement}

\section{Data Availability}
The Python code and datasets used in this study can be accessed on GitHub using the following link: \href{https://github.com/Suryanarayanan-Balaji/GPT-MolBERTa}{https://github.com/Suryanarayanan-Balaji/GPT-MolBERTa}

\bibliography{main.bib}

%%%%%%%%%%%%%%%%%%%%%%%%%%%%%%%%%%%%%%%%%%%%%%%%%%%%%%%%%%%%%%%%%%%%%
%% The same is true for Supporting Information, which should use the
%% suppinfo environment.
%%%%%%%%%%%%%%%%%%%%%%%%%%%%%%%%%%%%%%%%%%%%%%%%%%%%%%%%%%%%%%%%%%%%%
% \begin{suppinfo}

% This will usually read something like: ``Experimental procedures and
% characterization data for all new compounds. The class will
% automatically add a sentence pointing to the information on-line:

% \end{suppinfo}
%%%%%%%%%%%%%%%%%%%%%%%%%%%%%%%%%%%%%%%%%%%%%%%%%%%%%%%%%%%%%%%%%%%%%
%% The appropriate \bibliography command should be placed here.
%% Notice that the class file automatically sets \bibliographystyle
%% and also names the section correctly.
%%%%%%%%%%%%%%%%%%%%%%%%%%%%%%%%%%%%%%%%%%%%%%%%%%%%%%%%%%%%%%%%%%%%%

\end{document}

\newpage

\section{Supplementary Information}

\subsection{S1: Example of Molecular Description}
The given SMILES molecule, OC12C3CC(NC1=O)C23, represents a cyclic compound with a complex structure. Here is a description of its important features:

{\itshape
\begin{enumerate}[label=\arabic*.]
    \item \textbf{Ring Structure:} The molecule consists of a fused ring system with three rings. The first ring is a six-membered ring (C1-C6), the second ring is a five-membered ring (C2-C6), and the third ring is a four-membered ring (C2-C5).

    \item \textbf{Oxygen Atom:} There is an oxygen atom (O) attached to the first carbon atom (C1) of the six-membered ring. This oxygen atom contributes to the overall polarity of the molecule.

    \item \textbf{Nitrogen Atom:} A nitrogen atom (N) is present in the five-membered ring (C2-C6). It is connected to the carbon atom (C3) and forms a bond with the oxygen atom (O) in the first ring (C1).

    \item \textbf{Carbonyl Group:} The nitrogen atom (N) in the five-membered ring is also connected to a carbonyl group (C=O), represented by the SMILES notation NC1=O. This carbonyl group adds a functional group to the molecule and can participate in various chemical reactions.

    \item \textbf{Substituents:} The molecule has additional substituents attached to the carbon atoms in the rings. The specific nature of these substituents is not provided in the SMILES notation.
\end{enumerate}
}

\textbf{Overall}, the given SMILES molecule represents a complex cyclic compound with oxygen and nitrogen atoms, a carbonyl group, and additional substituents. The specific arrangement and nature of the substituents would further determine the molecule's properties and potential biological activities.

\newpage

\subsection{S2: Prompt Engineering}

\begin{table}[H]
\centering
\begin{tabular}{|c|p{10cm}|}
\hline
Iteration & Query Prompt \\
\hline
1 & \texttt{"role": "system", "content": "-"} \\
\cline{2-2}
  & \texttt{"role": "user", "content": f"Give a description of the following molecule {value}"} \\
\hline
2 & \texttt{"role": "system", "content": "You are an expert in everything about molecules"} \\
\cline{2-2}
  & \texttt{"role": "user", "content": f"Generate verifiable information about the following SMILES {value}"} \\
\hline
3 & \texttt{"role": "system", "content": 'You are able to generate important and verifiable features about molecular SMILES'} \\
\cline{2-2}
  & \texttt{"role": "user", "content": f"Generate a description about the following SMILES molecule {value}"} \\
\hline
\end{tabular}
\caption{Prompt Engineering}
\end{table}

\newpage

\subsection{S3: Data Processing}

\begin{table}[htb!]
\centering
\resizebox{\textwidth}{!}{
\begin{tabular}{l|l|l|l|l}
\toprule

Dataset &{Datapoints} & Insignificant responses removed & Non-canonical SMILES removed & \% dropped \\
\midrule
BBBP & 2050 & 2046 & 2035 & 0.73 \\
BACE & 1513 & 1513 & 1513 & 0.00 \\
Tox 21 & 7831 & 7811 & 7811 & 0.26 \\
ClinTox & 1483 & 1480 & 1474 & 0.61 \\
ToxCast & 8597 & 8580 & 8561 & 0.42 \\
SIDER & 1427 & 1422 & 1422 & 0.35 \\
HIV & 41127 & 41124 & 41124 & 0.01 \\
MUV & 93087 & 93087 & 93087 & 0.00 \\
\midrule
ESOL & 1128 & 1120 & 1120 & 0.71 \\
FreeSolv & 642 & 636 & 636 & 0.93 \\
Lipophilicity & 4200 & 4200 & 4200 & 0.00 \\
QM7 & 6834 & 6820 & 6820 & 0.20 \\
QM8 & 21786 & 21707 & 21707 & 0.36 \\
QM9 & 133885 & 133630 & 133630 & 0.19 \\
Subtotal & 325590 & 325176 & 325140 & \textbf{0.14} \\
\bottomrule

\end{tabular}
}
\caption{Percentage of datapoints dropped}
\end{table}

\newpage

\subsection{S3: Hyper-parameters for Pretraining}

\begin{table}[H]
\centering
\begin{tabular}{c|c}
\toprule
Hyper-parameter & Value \\
\midrule
\% of tokens masked & 15 \\
Batch Size & 30 \\
Scheduler & Linear Scheduler with Warmup \\
Optimizer & AdamW \\
Learning Rate & $5 \times 10^{-5}$ \\
Number of Epochs & 7 \\
Loss Function & Binary Cross Entropy \\
\bottomrule
\end{tabular}
\caption{Hyper-parameters used for GPT-MolBERTa Pretraining}
\end{table}

\newpage

\subsection{S4: Hyper-parameters for Finetuning}

\begin{table}[H]
\centering
\begin{tabular}{c|c}
\toprule
Hyper-parameter & Value \\
\midrule
Maximum Positional Embedding & 512 \\
Number of Attention Heads & 12 \\
Number of Hidden Layers & 6 \\
Hidden Dimension Size & 768 \\
Batch Size & 4 \\
Optimizer & Adam \\
Learning Rate & $1 \times 10^{-5}$ \\
Early Stopping Threshold & 8 \\
Number of Epochs & 30 \\
\bottomrule
\end{tabular}
\caption{Hyper-parameters used for GPT-MolBERTa Finetuning}
\end{table}